\documentclass[journal=nalefd,manuscript=letter,layout=twocolumn]{achemso}
\usepackage{chemformula} 
\usepackage[T1]{fontenc} 
\usepackage{pdfpages}

\author{Huilai Zhang}
\affiliation{Key Laboratory of Low-Dimensional Quantum Structures and Quantum Control of Ministry of Education, Department of Physics and Synergetic Innovation Center for Quantum Effects and Applications, Hunan Normal University, Changsha 410081, China}

\author{Ran Huang}
\affiliation{Key Laboratory of Low-Dimensional Quantum Structures and Quantum Control of Ministry of Education, Department of Physics and Synergetic Innovation Center for Quantum Effects and Applications, Hunan Normal University, Changsha 410081, China}

\author{Sheng-Dian Zhang}
\affiliation{Key Laboratory of Low-Dimensional Quantum Structures and Quantum Control of Ministry of Education, Department of Physics and Synergetic Innovation Center for Quantum Effects and Applications, Hunan Normal University, Changsha 410081, China}

\author{Ying Li}
\affiliation{Interdisciplinary Center for Quantum Information, State Key Laboratory of Modern Optical Instrumentation, College of Information Science and Electronic Engineering, Zhejiang University, Hangzhou 310027, China}
\alsoaffiliation{ZJU-Hangzhou Global Science and Technology Innovation Center, Key Lab. of Advanced Micro/Nano Electronic Devices $\&$ Smart Systems of Zhejiang, Zhejiang University, Hangzhou 310027, China}

\author{Cheng-Wei Qiu}
\affiliation{Department of Electrical and Computer Engineering, National University of Singapore, Singapore 117583, Singapore}

\author{Franco Nori}
\affiliation{Theoretical Quantum Physics Laboratory, RIKEN Cluster for Pioneering Research, Wako-shi, Saitama 351-0198, Japan}
\alsoaffiliation{Physics Department, The University of Michigan, Ann Arbor, Michigan 48109-1040, USA}

\author{Hui Jing}
\email{jinghui73@foxmail.com}
\affiliation{Key Laboratory of Low-Dimensional Quantum Structures and Quantum Control of Ministry of Education, Department of Physics and Synergetic Innovation Center for Quantum Effects and Applications, Hunan Normal University, Changsha 410081, China}

\title{Breaking Anti-$\mathcal{PT}$ Symmetry by Spinning a Resonator}
\abbreviations{IR,NMR,UV}
\keywords{Non-Hermitian physics, anti-$\mathcal{PT}$ symmetry, spinning resonator, nonreciprocal light manipulation, nanoparticle sensing}

\begin{document}

%

\begin{abstract}
	Non-Hermitian systems, with symmetric or antisymmetric Hamiltonians under the parity-time ($\mathcal{PT}$) operations, can have entirely real eigenvalues. This fact has led to surprising discoveries such as loss-induced lasing and topological energy transfer. A merit of anti-$\mathcal{PT}$ systems is free of gain, but in recent efforts on making anti-$\mathcal{PT}$ devices, nonlinearity is still required. Here, counterintuitively, we show how to achieve anti-$\mathcal{PT}$ symmetry and its spontaneous breaking in a linear device by spinning a lossy resonator. Compared with a Hermitian spinning device, significantly enhanced optical isolation and ultrasensitive nanoparticle sensing are achievable in the anti-$\mathcal{PT}$-broken phase. In a broader view, our work provides a new tool to study anti-$\mathcal{PT}$ physics, with such a wide range of applications as anti-$\mathcal{PT}$ lasers, anti-$\mathcal{PT}$ gyroscopes, and anti-$\mathcal{PT}$ topological photonics or optomechanics.
\end{abstract}

\section{Introduction}
Parity-time ($\mathcal{PT}$) symmetry provides a way to relax the conventional Hermiticity condition to ensure real eigenvalues of quantum systems~\cite{bender1998Real,ozdemir2019Parity}. Particularly, by steering the system parameters to surpass or encircle the  spectral degeneracy known as an exceptional point (EP)~\cite{miri2019Exceptional}, striking differences emerge in the properties of $\mathcal{PT}$ devices. These differences, confirmed in diverse systems with the gain-loss balance~\cite{guo2009Observation,ruter2010Observation,chang2014Parity,peng2014Parity,feng2014Singlemode,hodaei2014Paritytimesymmetric,wimmer2015Observation,li2019observation,zhu2014PT,zhu2018Simultaneous,schindler2011Experimental,bittner2012PT,assawaworrarit2017Robust,lu2015PTSymmetryBreaking,chen2018Generalized,dong2019Sensitive}, have created opportunities to achieve exotic functionalities, such as loss-induced lasing or anti-lasing~\cite{ozdemir2019Parity,feng2014Singlemode,hodaei2014Paritytimesymmetric}, robust wireless power transfer~\cite{assawaworrarit2017Robust}, and enhanced sensor responses~\cite{hodaei2017Enhanced,chen2018Generalized,dong2019Sensitive,xiao2019Enhanced} or light-matter interactions~\cite{jing2014PT,zhang2018Phonon,xu2016Topological}.

A tremendous effort has also been witnessed in achieving anti-$\mathcal{PT}$ symmetry which does not need any gain~\cite{ge2013antisymmetric}, thus providing a practical way to study quantum non-Hermitian effects~\cite{cao2020ReservoirMediated,wu2019Observation,klauck2019Observation}. Anti-$\mathcal{PT}$ symmetry has been demonstrated by using dissipatively coupled atomic beams~\cite{peng2016Antiparity}, and then by using cold atoms~\cite{jiang2019AntiParityTime}, electrical circuits~\cite{choi2018Observation}, thermal materials~\cite{li2019Paritytime}, and optical devices~\cite{zhang2019Dynamically,li2019Experimental,zhao2020Observation}. These breakthroughs have initiated the field of exploring unique anti-$\mathcal{PT}$ effects, e.g., energy-difference conserving dynamics~\cite{choi2018Observation} and chiral mode switching~\cite{zhang2019Dynamically}.

Here, we show how to achieve and utilize optical anti-$\mathcal{PT}$ symmetry breaking by spinning a linear resonator~\cite{maayani2018Flying}. We find that anti-$\mathcal{PT}$ symmetry emerging in such a device results in giant optical nonreciprocity and enhanced sensor performance. Our work is essentially different from the very recent work on anti-$\mathcal{PT}$ mode splitting in a nonlinear Brillouin device~\cite{zhang2020synthetic}, since: (i) the momentum or phase matching condition, as required for Brillouin devices, is not needed here; (ii) the power condition, i.e., above a threshold to create the Brillouin scattering, while not strong enough to induce other nonlinearities, is also unnecessary here; in fact, for a spinning resonator, the input can be ranging from single photons to a high-power laser~\cite{maayani2018Flying}; (iii) the spinning speed is much easier to be continuously tuned in situ than internal nonlinearity of materials. In a broader view, our work is well compatible with all the other well-established nonlinear or topological techniques, providing a feasible new way to engineer non-Hermitian devices, for applications in, e.g., anti-$\mathcal{PT}$ metrology~\cite{lai2019Observation,hokmabadi2019NonHermitian} and anti-$\mathcal{PT}$ topological photonics or nanomechanics~\cite{zhang2019Dynamically,lee2014Entanglement}.

\section{Results and discussions}
\textbf{Optical anti-$\mathcal{PT}$ symmetry.} Anti-$\mathcal{PT}$ systems with two coupled modes can be described at the simplest level as~\cite{ge2013antisymmetric}
\begin{equation}\label{Eq:Anti-PT_General_Form}
H_{\mathrm{APT}}=\left(\begin{array}{cc}
\omega & \kappa\\
-\kappa^{\ast} & -\omega^{\ast}
\end{array}\right),
\end{equation}
where $\omega$ is the complex frequency and $\kappa$ is the complex coupling strength between the two optical modes. It indicates two conditions are required for realizing anti-$\mathcal{PT}$ symmetry: (i) two excited modes with opposite frequency detunings and same loss or gain; (ii) anti-Hermitian coupling between the modes. In previous experiments, optical anti-$\mathcal{PT}$ symmetry has been realized based on e.g.,  complex spatial structures~\cite{peng2016Antiparity,zhang2019Dynamically} or nonlinear processes~\cite{jiang2019AntiParityTime,zhang2020synthetic}.
\begin{figure*}[ht]
	\includegraphics[width=0.92 \textwidth]{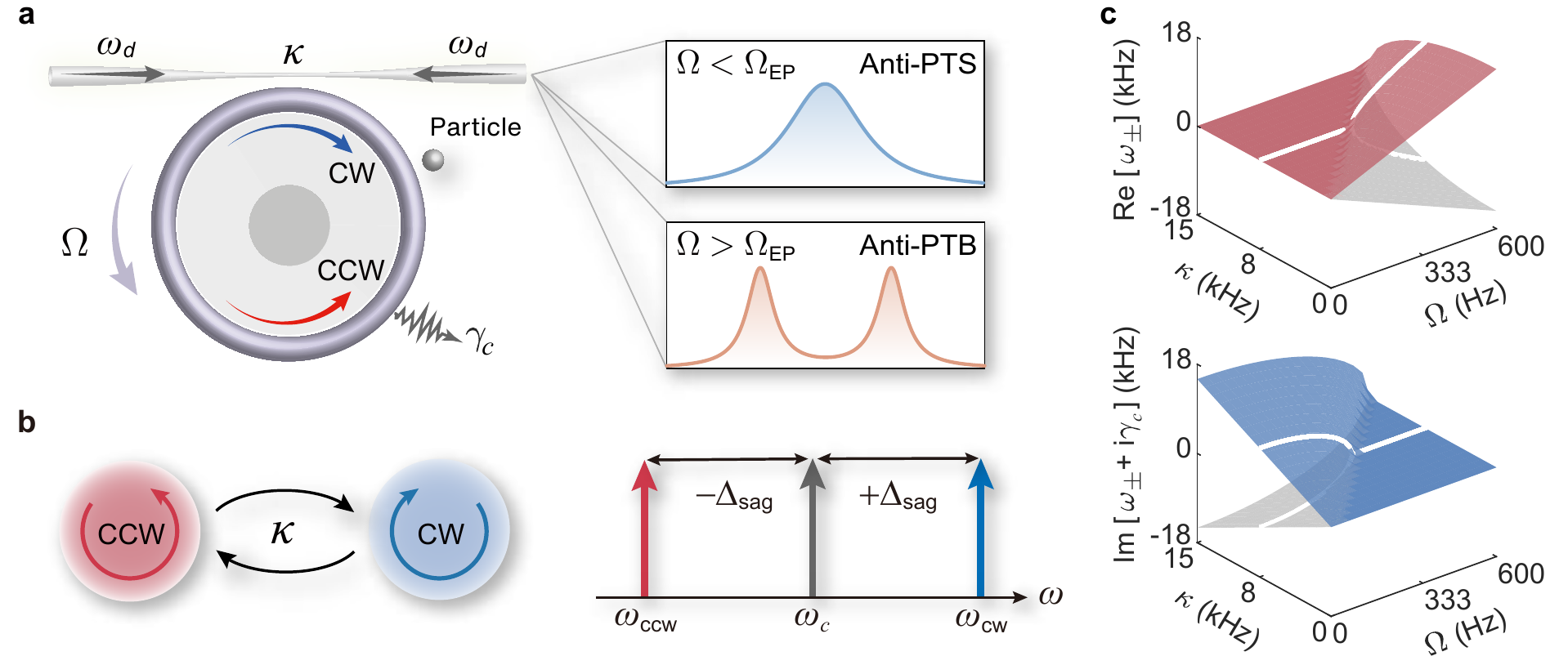}
	\caption{\label{Fig1}Anti-$\mathcal{PT}$ symmetry in a linear spinning resonator. (a) The resonator is driven by two lasers with the same frequency $\omega_{d}$ from the left and right. Anti-$\mathcal{PT}$ symmetry is realized with the opposite frequency shifts induced by the mechanical rotation with angular speed $\Omega$ and the dissipative coupling $\kappa$ induced by the taper scattering. The output spectra in the anti-$\mathcal{PT}$-symmetric (anti-PTS) and symmetry-broken (anti-PTB) phases can be observed by tuning $\Omega$ for $\Omega < \Omega_{\mathrm{EP}}$ and $\Omega > \Omega_{\mathrm{EP}}$, respectively, where $\Omega_{\mathrm{EP}}$ is the angular speed at EP. The nanoparticle is to be measured. (b) The schematics of Sagnac effect and dissipative coupling show the physical mechanism of realizing anti-$\mathcal{PT}$ symmetry. (c) The real and imaginary parts of eigenfrequencies versus $\kappa$ and $\Omega$ reveal the spectral properties of anti-$\mathcal{PT}$ symmetry. The white solid curves  correspond to the case of $\kappa=8\,\mathrm{kHz}$. For other parameter values, see the main text.}
\end{figure*}

\begin{figure}[h]
	\includegraphics[width=0.48 \textwidth]{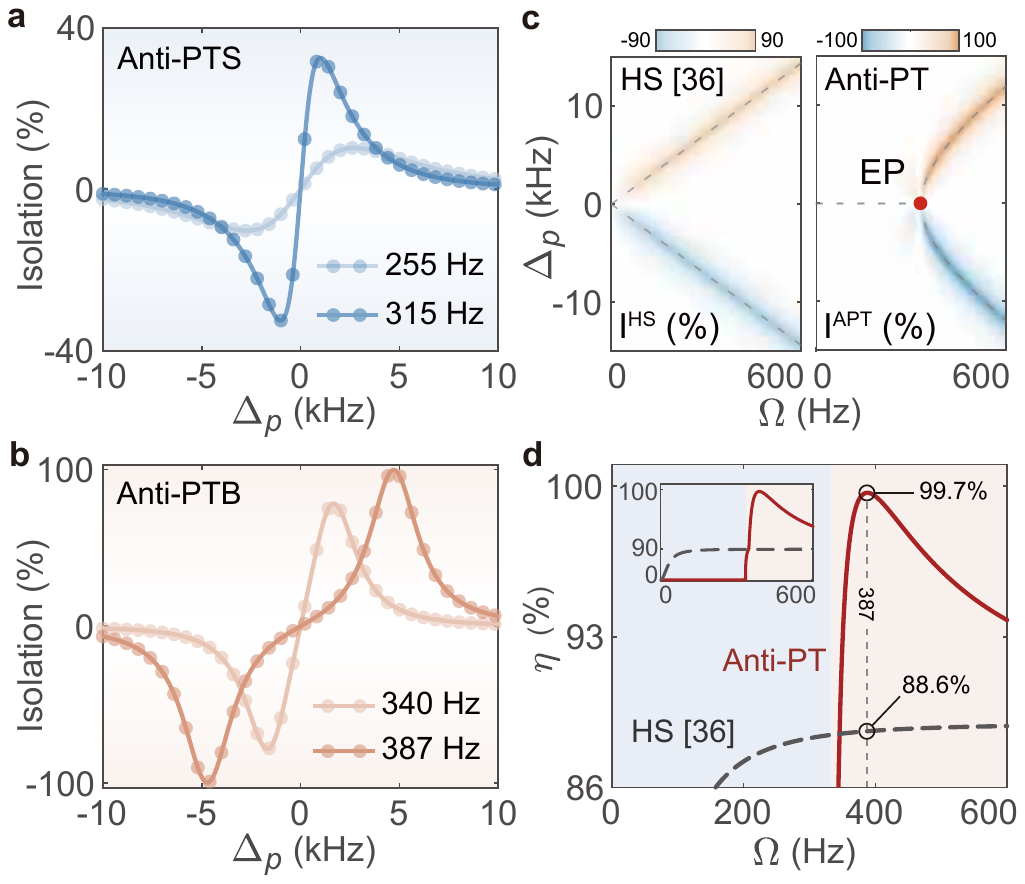}
	\caption{\label{Fig2} Nonreciprocal light transmission in the anti-$\mathcal{PT}$-symmetric system. (a) Isolation $I^{\mathrm{APT}}$ versus probe detuning $\Delta_p$ with different rotation speed $\Omega$ in the anti-$\mathcal{PT}$-symmetric phase and (b) symmetry-broken phase. (c) Isolations of Hermitian spinning (HS) resonator (left panel)~\cite{maayani2018Flying} and anti-$\mathcal{PT}$ system (right panel) versus $\Omega$ and $\Delta_p$. The gray dashed curves show the eigenfrequency evolutions of the two systems with respect to $\Omega$. (d) The selected maximum isolation $\eta$ as a function of $\Omega$ for anti-$\mathcal{PT}$ (red solid line) and HS (dark dashed line) systems. The other parameters are the same as those in Figure~\ref{Fig1}.}
\end{figure}

As already demonstrated in the experiment~\cite{lai2019Observation}, see Figure~\ref{Fig1}a, we consider a linear optical resonator driven by two lasers with frequency $\omega_{d}$ from the left and right, which can excite the clockwise (CW) and counter-clockwise (CCW) travelling modes. When the resonator is spinning at an angular velocity $\Omega$~\cite{maayani2018Flying}, the rotation-induced Sagnac–Fizeau shift $\omega_c\to\omega_c\pm\Delta_\mathrm{sag}$ is given by~\cite{malykin2000Sagnac}
\begin{equation}\label{Eq:Delta_sag}
\Delta_{\text{sag}}=\frac{nR\Omega\omega_{c}}{c}\left(1-\frac{1}{n^{2}}-\frac{\lambda}{n}\frac{\mathrm{d}n}{\mathrm{d}\lambda}\right),
\end{equation}
where $\omega_{c}$ is the resonant frequency of a nonspinning resonator, $c$ ($\lambda$) is the speed (wavelength) of light, $n$ and $R$ are the refractive index and radius of the resonator, respectively. The dispersion term $\mathrm{d}n/\mathrm{d}\lambda$, characterizing the relativistic origin of the Sagnac effect, is relatively small in typical materials ($\sim1\!\%$)~\cite{maayani2018Flying}. We fix the CCW rotation of the resonator; hence $+\Delta_\mathrm{sag}$ ($-\Delta_\mathrm{sag}$) corresponds to the CW (CCW) travelling mode, see Figure~\ref{Fig1}b. The taper-scattering-induced dissipative backscattering leads to the dissipative coupling $i\kappa$ between the countercirculating modes~\cite{lai2019Observation}, which is anti-Hermitian, i.e., $-(i\kappa)^{\ast}=i\kappa$. Obviously, this Sagnac resonator naturally fulfills the conditions of the realization of anti-$\mathcal{PT}$ symmetry.

In a frame rotating at driving frequency $\omega_d$, the effective Hamiltonian of this system is
\begin{equation}\label{Eq:Hamiltonian}
H_0=\left(\begin{array}{cc}
\Delta_{+}-i\gamma_c & i\kappa\\
i\kappa & \Delta_{-}-i\gamma_c
\end{array}\right),
\end{equation}
where $\Delta_{\pm}=\Delta_{c} \pm \Delta_{\mathrm{sag}}$ are the optical detunings in the spinning case with the optical driving detuning $\Delta_{c}=\omega_{c}-\omega_{d}$, and $\gamma_c = \left(\gamma_0+\gamma_\mathrm{ex}\right)/2$ is the total optical loss including the intrinsic loss of the resonator $\gamma_0 \equiv \omega_c/Q$ with the quality factor $Q$ and the loss due to the coupling of the resonator with the fiber taper $\gamma_{\mathrm{ex}}$. For $\Delta_{c} = 0$, anti-$\mathcal{PT}$ symmetry can be realized without any gain or nonlinearity, which is different from that using nonlinear Brillouin scattering to provide optical gain in a resonator~\cite{zhang2020synthetic}. Unlike $\mathcal{PT}$ symmetry, anti-$\mathcal{PT}$ symmetry is independent on spatially separated gain-loss balanced structure.

The eigenfrequencies of this linear anti-$\mathcal{PT}$-symmetric system are
\begin{align}\label{Eq:eigenfrequencies}
	\omega_{\pm} & =-i\gamma_{c}\pm\sqrt{\Delta_{\mathrm{sag}}^{2}-\kappa^{2}},
\end{align}
indicating a phase transition as $\Omega$ varies. Figure~\ref{Fig1}c shows that, for small $\Omega$ ($\Delta_{\mathrm{sag}}<\kappa$), the eigenmodes preserve anti-$\mathcal{PT}$ symmetry with the same resonance frequency but different linewidths. The symmetry breaking occurs at EP ($\Delta_{\mathrm{sag}}=\kappa$) where the eigenstates coalesce. For large $\Omega$ ($\Delta_{\mathrm{sag}}>\kappa$), the system enters the symmetry-broken phase with bifurcating eigenmodes.
For a  specific dissipative coupling strength, e.g., $\kappa = 8\,\mathrm{kHz}$, the critical value of rotation speed is obtained as $\Omega_{\mathrm{EP}}=333\,\mathrm{Hz}$ (see the white solid curves in Figure~\ref{Fig1}c).

Here, we take the experimentally accessible parameters~\cite{armani2003Ultrahigh,huet2016Millisecond,peng2014Lossinduced}: $\lambda = 1550\ \mathrm{nm}$, $Q \approx 1\times10^{11}$, $\gamma_{\mathrm{ex}} = \gamma_0/2$, $n = 1.44$, and $R = 50 \ \mu\mathrm{m}$. As demonstrated in the experiment~\cite{lai2019Observation}, the dissipative coupling originating from taper scattering can be $\sim 8\,\mathrm{kHz}$.
In a recent experiment~\cite{maayani2018Flying}, light in a tapered fiber is evanescently coupled into or out of a linear resonator (mounted on a spinning turbine). The spinning can drag air into the coupling region, so that a boundary layer of air forms. Due to the air pressure, the taper lies at a height above the resonator, which can be several nanometers. If perturbation induces the taper to rise higher than the equilibrium, it floats back to its original position, a process called aerodynamical self-adjustment~\cite{maayani2018Flying}. This process thus leads to a stable coupling between the fiber and the resonator, and the taper does not touch or stick to the spinning device even if the taper is pushed towards it~\cite{maayani2018Flying}. This device is robust against the extra loss induced by the internal defects or other dissipative elements in practice. Also, as demonstrated in an experiment~\cite{qin2020Fast}, the Fizeau shift can be increased even for a slow rotary speed, by using a dispersion enhanced technique~\cite{qin2020Fast}.

\textbf{Symmetry-broken nonreciprocity.} When a probe light of frequency $\omega_{p}$ is incident from the left (right) side and using the input-output relation (see the Supporting Information), the transmission rate $T_\mathrm{cw}$ ($T_\mathrm{ccw}$) is 
\begin{align}\label{Eq:T_APT_HS}
	T_{\mathrm{cw,ccw}}\!=\!\left|1\!+\!\frac{i\gamma_{\mathrm{ex}}\left(\delta_{p}\mp\Delta_{\mathrm{sag}}\right)}{\left(\delta_{p}\!+\!\Delta_{\mathrm{sag}}\right)\left(\delta_{p}\!-\!\Delta_{\mathrm{sag}}\right)\!+\!\kappa^{2}}\right|^{2},
\end{align}
where $\delta_{p} = \Delta_{p}-i\gamma_{c}$, $\Delta_p=\omega_c-\omega_p$. Clearly, the term $\mp\Delta_{\mathrm{sag}}$ is the origin of nonreciprocal light transmission. In the anti-$\mathcal{PT}$-symmetric phase, $\Delta_{\mathrm{sag}}<\kappa$, the difference of the two transmission rates is limited, while in the symmetry-broken phase, this difference becomes larger for $\Delta_{\mathrm{sag}}>\kappa$, enabling better one-way transmission. To confirm this picture, we study the isolation of this anti-$\mathcal{PT}$ system, i.e.,
$I^\mathrm{APT}=\left|T_{\mathrm{cw}} - T_{\mathrm{ccw}}\right|$, with normalized $T_\mathrm{cw,ccw}$. Indeed, the symmetry-broken isolation is much larger than that in the symmetric phase, see Figures~\ref{Fig2}a and \ref{Fig2}b.

For a comparison, the optical isolation by using a Hermitian spinning (HS) resonator $I^\mathrm{HS}$~\cite{maayani2018Flying} and the selected maxima of the isolation $\eta\equiv\max\,[I]$ are also plotted in Figures~\ref{Fig2}c and \ref{Fig2}d, respectively. The isolation of HS system becomes larger by increasing $\Omega$ due to the optical mode splitting induced by the Sagnac shift; but it is limited to $90\%$ for the parameter values used here, due to the fixed linewidth~\cite{maayani2018Flying}. In contrast, for an anti-$\mathcal{PT}$ device, both the linewidths and the transmission can be altered, resulting in an enhanced isolation as high as $99.7\%$. This anti-$\mathcal{PT}$-broken nonreciprocity, due to the interplay of synthetic angular momentum and dissipative backscattering, is fundamentally different from $\mathcal{PT}$-symmetric nonreciprocity originating from nonlinearities~\cite{chang2014Parity,peng2014Parity}. One-way devices, free of exquisite control of gain or nonlinearity, have such a wide range of applications as on-chip circulator~\cite{peng2014Parity,chang2014Parity}, invisible sensing~\cite{fleury2015Invisible,cai2007Optical}, and quantum optical computation~\cite{knill2001Scheme}.

\textbf{Anti-$\mathcal{PT}$ sensor.} In the presence of the external perturbation, the variation of the transmission spectrum also gives a way to detect the perturbation itself. For an example, taking into account of a single nanoparticle falling into or flying by the evanescent field of the resonator~\cite{jing2018Nanoparticle}, see Figure~\ref{Fig1}a, we have the modified Hamiltonian of the perturbated system~\cite{chen2017Exceptional,jing2018Nanoparticle}
\begin{figure*}[h]
	\includegraphics[width=0.97 \textwidth]{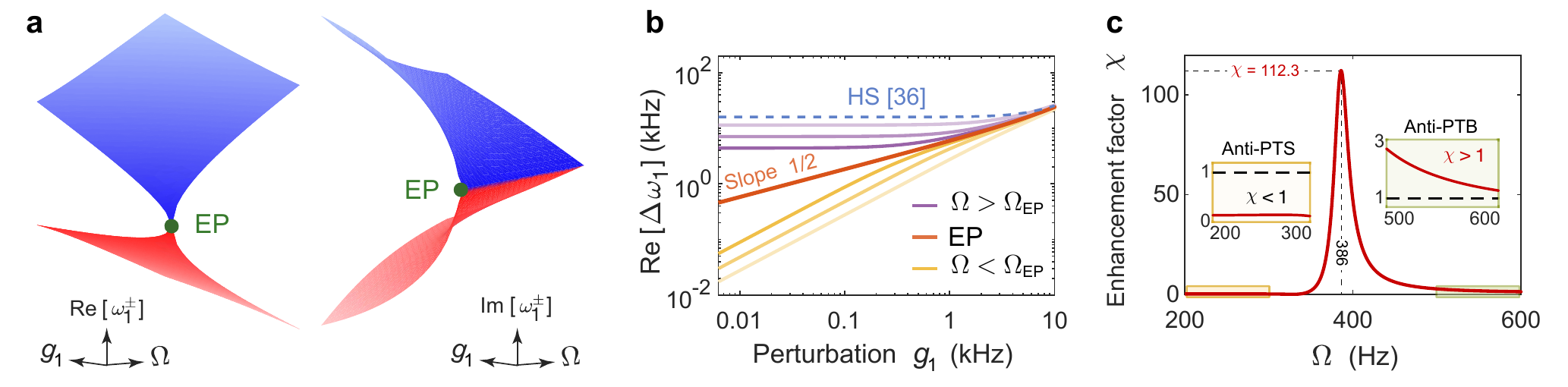}
	\caption{\label{Fig3} Anti-$\mathcal{PT}$-symmetric ultrasensitive nanoparticle sensing. (a) The evolution of the eigenenergy of the perturbated system versus the perturbation $g_1$ induced by a single nanoparticle. Here we set $\gamma_1 / g_1 = 0.05$ as that in the experiments~\cite{chen2017Exceptional,peng2016Chiral}. (b) Dependence of frequency splitting $\mathrm{Re}\,[\Delta\omega_{1}]$ corresponding to anti-$\mathcal{PT}$ sensor (solid curves) and HS sensor (dashed line)~\cite{maayani2018Flying} on $g_1$. The yellow, purple and orange curves denote the cases in anti-$\mathcal{PT}$-symmetric ($\Omega < \Omega_{\mathrm{EP}}$) or symmetry-broken phase ($\Omega > \Omega_{\mathrm{EP}}$), and at EP ($\Omega = \Omega_{\mathrm{EP}}$), respectively. The rotation speed of HS sensor is chosen as $\Omega = \Omega_{\mathrm{EP}}=333 \,\mathrm{Hz}$. (c) Enhancement factor $\chi$ as a function of $\Omega$. Here, $g_1=3\,\mathrm{kHz}$ and the probe detuning $\Delta_p$ is set to $\mathrm{Re}\,[\omega_{-}]$. The other parameters are the same as those in Figure~\ref{Fig1}.}
\end{figure*}
\begin{equation}\label{Eq:Perturbed_Hamiltonian}
H_{1}=\left(\begin{array}{cc}
\Delta_{+}^{\prime}-i\gamma_{c}+J & i\kappa+J\\
i\kappa+J & \Delta_{-}^{\prime}-i\gamma_{c}+J
\end{array}\right),
\end{equation}
where $\Delta^{\prime}_{\pm} = \Delta_p \pm \Delta_{\mathrm{sag}}$, and $J=g_{1}-i\gamma_{1}$ is the complex perturbation induced by the nanoparticle with frequency shift  $g_1$ and the linewidth broadening $\gamma_1$. As shown in Figure~\ref{Fig3}a, the eigenfrequency structure reveals how this system reacts on a sufficiently small perturbation. The sensitivity can be defined as the difference between the two eigenfrequencies, i.e.,
\begin{equation}\label{Eq:Perturbed_frequency_splitting1}
\Delta\omega_{1}=2\sqrt{\Delta_{\mathrm{sag}}^{2}-\left(\kappa-iJ\right)^{2}}.
\end{equation}

Figure \ref{Fig3}b, showing the logarithmic behaviour of $\mathrm{Re}\,[\Delta\omega_{1}]$, highlights the sensitivity enhancement of anti-$\mathcal{PT}$ sensor, compared to that using an HS device~\cite{maayani2018Flying,jing2018Nanoparticle}. For the same perturbation, HS sensor performs closely to anti-$\mathcal{PT}$ sensor when operating in the symmetry-broken phase. In the anti-$\mathcal{PT}$-symmetric phase, the splitting is smaller than that in symmetry-broken phase. However, at EP, the slope of the response is $1/2$, which can be explained by using perturbation theory. When $J$ is much smaller than $\kappa$, the complex frequency splitting is expected to approximately follow
\begin{equation}\label{Eq:Perturbed_frequency_splitting2}
\Delta\omega_{1}=2\sqrt{2i\kappa} J^{1/2}.
\end{equation}
For larger $J$, the slope is slightly larger than $1/2$ because in this case eq~\ref{Eq:Perturbed_frequency_splitting1} cannot be simplified to eq~\ref{Eq:Perturbed_frequency_splitting2} (see more details in the Supporting Information). In experiments~\cite{chen2017Exceptional,hodaei2017Enhanced}, the sensitivity defined by frequency splitting can be assessed by monitoring the separation of the spectral lines in transmission spectrum.

Sensitive responses to perturbations can also be revealed by measuring the variation of transmission spectrum~\cite{jing2018Nanoparticle}. Choosing, e.g., the CW mode, the transmittance variation can be defined as $\mathcal{V}_{s} \equiv T^{s}_{1}/T^{s}_{0}$, where $s$ denotes the anti-$\mathcal{PT}$ or HS sensor and $T^{s}_{1,0}$ is the transmission with or without the perturbation. Then we define the factor
\begin{align}
	\chi & = \mathcal{V}_{\mathrm{APT}}/\mathcal{V}_{\mathrm{HS}},
\end{align}
to show the sensitivity enhancement of the anti-$\mathcal{PT}$ sensor. We find that by breaking the anti-$\mathcal{PT}$ symmetry, $\chi > 1$ can be achieved. In particular, for $\Omega \sim 386\,\mathrm{Hz}$, i.e., when the isolation reaches its maximum, 2 orders of magnitude enhanced sensitivity is achievable (see Figure~\ref{Fig3}c). This provides a conceptually new way to engineer an optical resonator to realize ultrasensitive nanoparticle sensors, as crucial elements in medical diagnosis and environmental monitoring~\cite{vollmer2008Single,zhu2010Onchip,chen2017Inflection}. In a broader view, by extending to a spinning nonlinear or optomechanical resonator, anti-$\mathcal{PT}$ devices can also be used to enhance the responses of e.g., gyroscopes or weak-force sensors.

In experiments~\cite{peng2016Chiral,zhu2010Onchip}, nanoparticle sensing has been realized by measuring mode splitting. The particle falling through the mode volume of the resonator has a limited interaction time with the field, and, regardless of whether this time is very short or long, the resonator will feel and respond to it by exhibiting mode splitting~\cite{zhu2010Controlled}. A spinning resonator may induce the particle to diffuse in the air, which may change the overlap of the particle and the mode volume. The values of the frequency shift $g_{1}$ and the linewidth broadening $\gamma_{1}$ are thus changed by the particle position, but their ratio $\gamma_{1}/g_{1}$ remains unaffected (see the Supporting Information for details). Also, the particle position does not affect the Sagnac effect, and its detection in current experimental whispering-gallery-mode sensors can be very fast~\cite{peng2016Chiral,zhu2010Onchip}. In the future, we will study more complicated cases such as the impact of time-varying coupling of the resonator and the flying-by nanoparticles, and improved sensing efficiency by using, e.g., the dispersion enhanced technique~\cite{qin2020Fast}.

\section{Conclusions and outlook} We have revealed anti-$\mathcal{PT}$ symmetry in a spinning resonator, without any optical nonlinearity or complex spatial structures. The opposite Sagnac frequency shifts and the optical dissipative coupling enable the system Hamiltonian to be antisymmetric under the combined $\mathcal{PT}$ operations. In particular, by breaking the anti-$\mathcal{PT}$ symmetry, enhanced optical nonreciprocity and nanoparticle sensing can be achieved in this linear device. Our work provides a highly feasible way to achieve anti-$\mathcal{PT}$ symmetry breaking and can be further extended to study nonlinear or higher order anti-$\mathcal{PT}$ effects by spinning a nonlinear device~\cite{cao2019HighOrder}.

Our scheme will not, of course, render other existing techniques obsolete. On the contrary, an anti-$\mathcal{PT}$ isolator can provide one-way light flow, while nonlinearity or synthetic dimensions can be integrated to induce or engineer nonlinear or quantum effects in anti-$\mathcal{PT}$ devices~\cite{wu2019Observation,cao2020ReservoirMediated}. Our anti-$\mathcal{PT}$ resonator is thus well compatible with other well-developed techniques~\cite{zhang2019Dynamically,zhang2020synthetic}, indicating exciting possibilities in non-Hermitian optics or nanomechanics~\cite{xu2016Topological,yuan2018Synthetic}.

\bigskip

\textbf{Note added}: After the submission of our manuscript, a new paper was posted on arXiv, reporting experimental observations of anti-$\mathcal{PT}$ topology and anti-$\mathcal{PT}$-enhanced Brillouin sensing in a fiber (with nonlinearity and gain)~\cite{bergman2020Observation}.

\begin{acknowledgement}
We thank Li Ge at the City University of New York for his constructive suggestions. H.J. is supported by the National Natural Science Foundation of China under Grants No. 11935006 and No. 11774086. C.-W.Q. acknowledges the financial support from A*STAR Pharos Program (Grant No. 15270 00014, with Project No. R-263-000-B91-305) and Ministry of Education, Singapore (Project No. R-263-000-D11-114). F.N. is supported in part by: NTT Research, Army Research Office (ARO) (Grant No. W911NF-18-1-0358), Japan Science and Technology Agency (JST) (via the Q-LEAP program and the CREST Grant No. JPMJCR1676), Japan Society for the Promotion of Science (JSPS) (via the KAKENHI Grant No. JP20H00134, and the JSPS-RFBR Grant No. JPJSBP120194828), and the Foundational Questions Institute Fund (FQXi) (Grant No. FQXi-IAF19-06), a donor advised fund of the Silicon Valley Community Foundation.
\end{acknowledgement}

\begin{suppinfo}
The following files are available free of charge.
\begin{itemize}
  \item Experimental feasibility analysis and detailed derivations of the optical spectrum.
\end{itemize}

\end{suppinfo}


\providecommand{\latin}[1]{#1}
\makeatletter
\providecommand{\doi}
{\begingroup\let\do\@makeother\dospecials
	\catcode`\{=1 \catcode`\}=2 \doi@aux}
\providecommand{\doi@aux}[1]{\endgroup\texttt{#1}}
\makeatother
\providecommand*\mcitethebibliography{\thebibliography}
\csname @ifundefined\endcsname{endmcitethebibliography}
{\let\endmcitethebibliography\endthebibliography}{}

\clearpage


\section*{Supplementary material}

# Supporting Information for

## Breaking Anti-$\mathcal{PT}$ Symmetry by Spinning a Resonator

Huilai Zhang,[†] Ran Huang,[†] Sheng-Dian Zhang,[†] Ying Li,[‡,¶] Cheng-Wei Qiu,[§] Franco Nori,[∥,⊥] and Hui Jing[*,†]

[†]*Key Laboratory of Low-Dimensional Quantum Structures and Quantum Control of Ministry of Education, Department of Physics and Synergetic Innovation Center for Quantum Effects and Applications, Hunan Normal University, Changsha 410081, China*
[‡]*Interdisciplinary Center for Quantum Information, State Key Laboratory of Modern Optical Instrumentation, College of Information Science and Electronic Engineering, Zhejiang University, Hangzhou 310027, China*
[¶]*ZJU-Hangzhou Global Science and Technology Innovation Center, Key Lab. of Advanced Micro/Nano Electronic Devices & Smart Systems of Zhejiang, Zhejiang University, Hangzhou 310027, China*
[§]*Department of Electrical and Computer Engineering, National University of Singapore, Singapore 117583, Singapore*
[∥]*Theoretical Quantum Physics Laboratory, RIKEN Cluster for Pioneering Research, Wako-shi, Saitama 351-0198, Japan*
[⊥]*Physics Department, The University of Michigan, Ann Arbor, Michigan 48109-1040, USA*
∗ *jinghui73@foxmail.com*

Here, we present technical details on anti-parity-time ($\mathcal{PT}$)-symmetry-broken nonreciprocity in a linear resonator. In Sec. S1, we discuss the experimental feasibility of our scheme. In Sec. S2, we provide the detailed derivations of the anti-$\mathcal{PT}$-symmetric Hamiltonian. In Sec. S3, we show the extended results about nonreciprocal light transmission and ultrasensitive nanoparticle sensing.

## S1. EXPERIMENTAL FEASIBILITY

### A. Stable coupling between the tapered fiber and the spinning resonator

To realize anti-$\mathcal{PT}$ symmetry, one indispensable condition is two excited modes with opposite frequency detunings. Inspired by a recent experiment [S1], we find due to the Sagnac effect, counterpropagating modes with opposite frequency shifts naturally exist in a spinning resonator, which is pumped bidirectionally. Mounting a silica microtoroid resonator on a turbine and positioning it near a tapered region of a single-mode telecommunication fiber, the light can be coupled into or out of the spinning cavity evanescently.

#### 1. Self-adjustment process and critical coupling

According to the experiment [S1], the aerodynamic process plays a key role in stable resonator-fiber coupling. When the resonator rotates at an angular velocity $\Omega$, a boundary layer of air will be dragged into the region between the taper and the resonator. Hence, the taper will fly above the resonator with a separation of several nanometers. If some perturbations cause the taper to rise higher than the stable-equilibrium height, it floats back to its original position, which is called "self-adjustment".

Using similar methods in Ref. [S1], we consider the local deformation of the taper [see Fig. S1(a)], break it into a set of infinitesimal cylinders and focus on the outermost one. The air pressure on this infinitesimal cylinder, leading to a tiny displacement $d$, is written as $\Delta T_\mathrm{air} = (\rho \Delta \theta) \, T_\mathrm{air}/L$, where $\rho$ ($\theta$) represents the radius (angle) of the winding shape for the deformed region of the fiber. Total pressure on the taper from a boundary layer of air (the "air bearing" surface), $T_\mathrm{air}$, can be estimated analytically from [S1]

$$T_\mathrm{air} = 6.19 \mu R^{5/2} \Omega \int_0^r \left( h - \sqrt{r^2 - x^2} + r \right)^{-3/2} \mathrm{d}x, \tag{S1}$$

where $\mu$ is the viscosity of air, $r$ ($R$) is the radius of the taper (resonator), $h = h_0 + d$ is the taper-resonator separation with $h_0$ being the gap between the taper and the surface of the non-spinning resonator. Note that in this device, the curvature of the microtoroid is so small compared with the width of the film-lubricated region of interest, thus its surface can be considered as a flat plane, which is similar to the case of spinning sphere in Ref. [S1].

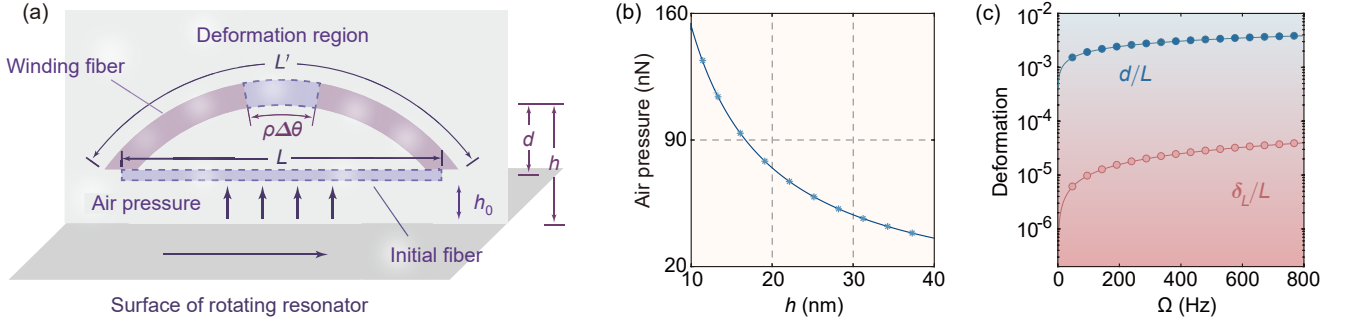

FIG. S1. (a) Schematics of the "self-adjustment" behaviour. (b) The air pressure versus fiber-resonator separation at $\Omega = 700\,\text{Hz}$. (c) The strain and the displacement as a function of the angular velocity for $h = 20\,\text{nm}$.

The tension on this infinitesimal cylinder induced by the deformation can be calculated by $\Delta T_{\text{ela}} = 2\mathcal{F}\sin(\Delta\theta/2) \approx \mathcal{F}\Delta\theta$, where the elastic force on the taper $\mathcal{F}$ obeys Hooke's law:

$$\sigma = E\epsilon. \tag{S2}$$

Here, E is the Young's modulus of silica, $\sigma = \mathcal{F}/(\pi r^2)$ represents the uniaxial stress, $\epsilon = \delta_L/L$ is the strain, $L$ stands for the original length of the deformation region of the taper, and $\delta_L = L' - L$ denotes the change in length, which can be derived with the aid of $L' = \rho\theta$, $(L/2)^2 + (\rho - d)^2 = \rho^2$, and $\sin(\theta/2) = L/(2\rho)$. Therefore, in the case of stable equilibrium ($\Delta T_{\text{air}} = \Delta T_{\text{ela}}$), we can describe $T_{\text{air}}$ in another form:

$$T_{\text{air}} = 2\pi r^2 E\left[\arcsin(\phi) - \phi\right] \approx \pi r^2 \phi^3 E/3, \tag{S3}$$

with $\phi = 4Ld/(L^2 + 4d^2)$ being much smaller than 1 under the approximation of $d/L \ll 1$. The inverse trigonometric function $\arcsin(\phi)$ can be expanded using a Taylor series: $\arcsin(\phi) = \phi + \phi^3/6 + \cdots$ for $|\phi| \ll 1$; then, the displacement $d$ caused by air pressure can be analytically estimated as

$$d = L\left(\tau - \sqrt{\tau^2 - 1}\right)/2, \tag{S4}$$

where $\tau = \left[\pi r^2 E/(3T_{\text{air}})\right]^{1/3}$. Accordingly, the strain of the taper can be rewritten as

$$\epsilon = \arcsin(\phi)/\phi - 1 \approx \phi^2/6. \tag{S5}$$

From this equation, we find the strain, i.e., the elastic force, is positively associated with the distance between the fiber and the surface of the rotating resonator:

$$\frac{\partial \mathcal{F}}{\partial h} = \pi r^2 E\left(\frac{\partial \epsilon}{\partial d}\right) = \frac{16\pi r^2 E L^2 d\left(L^2 - 4d^2\right)}{3\left(L^2 + 4d^2\right)^3} > 0. \tag{S6}$$

Therefore, if any perturbation causes the gap to be larger than the stable-equilibrium distance, the elastic force will be stronger.

The "self-adjustment" behavior can be understood from the responses of air pressure and elastic force to the variation of the gap induced by a perturbation. As shown in Fig. S1(b), the air pressure is reduced dramatically when the fiber is far away from the resonator. Meanwhile, the elastic force becomes larger due to the stronger deformation, which indicates that the fiber can be dragged back to the original position; thus we can maintain the separation between the spinning devices and the couplers, which is essential for critical coupling [S1].

Moreover, to realize the fast spinning resonator experimentally, it requires the microtoroid cavity to be perfectly circular to maintain a stable taper-resonator coupling. While the experimental toolbox and the fabrication methods of microresonators, in principle silica based whispering-gallery-mode microresonators in the form of toroid or sphere are mature. The resonators can be fabricated with almost atomically small surface roughness and close-to perfect shape via surface tension or polishing techniques. Thus, the impact of the resonator shape is negligible and the coupling with the waveguide can be maintained stably.

Herein, we choose experimentally accessible parameters [S1, S2]: $E = 75\,\text{GPa}$, $r = 544\,\text{nm}$ and $L = 5\,\mu\text{m}$. As presented in Fig. S1(c), the deformation is extremely small ($d/L < 0.37\%$ and $\epsilon < 3.625 \times 10^{-5}$), thereby our approximation is physically reasonable.

### 2. Intermolecular forces

The intermolecular forces between the taper and the spinning resonator, including Casimir and van der Waals forces, can be described as [S1]:

$$T_{\text{int}} = rR\left(-\frac{\mathbb{A}}{6\pi h^3} + \frac{\mathbb{B}}{45\pi h^9} - \frac{\pi^2 c\hbar}{240 h^4}\right), \tag{S7}$$

where $\mathbb{B}$ is a constant [S3], and $\mathbb{A}$ is the Hamaker constant, which can be calculated from the following equation [S4]:

$$\mathbb{A} = \frac{3\varepsilon_-^{(1)}\varepsilon_-^{(2)} k_{\text{B}} \text{T}}{4\varepsilon_+^{(1)}\varepsilon_+^{(2)}} + \frac{\nu\left[n_-^{(1)} n_-^{(2)}\right]^2}{n_+^{(1)} n_+^{(2)}\left[n_+^{(1)} + n_+^{(2)}\right]}, \tag{S8}$$

with $\nu = 3\sqrt{2}\hbar\nu_e/16$, $\varepsilon_\pm^{(j)} = \varepsilon_j \pm \varepsilon_0$, $n_\pm^{(j)} = \sqrt{n_j^2 \pm n_0^2}$, and $j = 1, 2$. Moreover, $\varepsilon_0$ ($n_0$), $\varepsilon_1$ ($n_1$) and $\varepsilon_2$ ($n_2$) represent the dielectric constant (the refractive index) of air, taper and spinning resonator, respectively; $\nu_e$ is a constant, $k_{\text{B}}$ is the Boltzmann constant, and T is the temperature of the system [S4].

In previous studies, it has been shown that the intermolecular forces begin to attract the flyer towards the rotor when the gap between them is reduced to less than 10 nm, and to strongly repel them when the gap is narrowed further, normally to below 300 fm [S5]. In our system, the fiber-resonator separation is set to be 20 nm, thus we can safely omit the effects of Casimir and van der Waals forces. The experiment [S1] also shows some other factors, such as lubricant compressibility, tapered-fiber stiffness, and wrap angle of the fiber, may affect resonator–waveguide coupling. However, the effects induced by these factors are confirmed to be negligible in the experiment.

### 3. Air friction

The air drag torque on the rotational resonator is given by [S6]:

$$M_{\text{air}} = 1.336 \Omega P_0 R^4 \sqrt{\frac{2\pi m_0}{k_{\text{B}} \text{T}_0}} \propto \Omega, \tag{S9}$$

where $P_0$ is the air pressure, $m_0 = 4.6 \times 10^{-26}$ kg is the mass of the air molecule, and $\text{T}_0$ is the temperature of the surrounding air molecules. We note that Eq. (S9) is originally applied to a single spinning sphere. However, we can use it here since the surface of the microtoroid resonator can also be regarded as a flat plane within the film-lubricated region, as presented in Sec. S1 A 1. Moreover, the rotation speed used in our numerical simulations is below kHz level, which is comparatively small. For this reason, we can neglect rotation-induced heating on the air molecules, and assume $\text{T}_0 = \text{T} = 300$ K and $P_0 = 1.013$ bar. In our approximations, the maximum value of air drag torque is $M_{\text{air}} = 4.947 \times 10^{-12}$ N · m, which indicates air friction can be safely ignored in our system.

### 4. Stability analysis

The condition of $\tau \geq 1$ in Eq. (S4) yields the first limit of angular velocity:

$$\Omega_0 = \frac{\varrho \pi r^2 \text{E}}{18.57 \mu R^{5/2}}, \tag{S10}$$

where

$$\varrho = \left[\int_0^r \left(h - \sqrt{r^2 - x^2} + r\right)^{-3/2} \mathrm{d}x\right]^{-1}. \tag{S11}$$

Also, the approximation condition $d \ll L$ gives another limit ($d/L = 1\%$):

$$\Omega_1 = \frac{\varrho \pi r^2 \text{E}}{\upsilon \mu R^{5/2}} < \Omega_0, \tag{S12}$$

where $\upsilon = 2.905 \times 10^5$. Furthermore, the tiny displacement $d$ should be smaller than the taper-resonator separation $h$, thus provides the third limit:

$$\Omega_2 = \frac{\varrho \pi r^2 \Lambda \mathrm{E}}{18.57 \mu R^{5/2}}, \tag{S13}$$

with $\Lambda = \left[4Lh/\left(L^2 + 4h^2\right)\right]^3$. Finally, we consider the elastic limit of the material for the taper ($\sigma = \Upsilon$):

$$\Omega_3 = \frac{\varrho \pi r^2 \Upsilon}{3.095 \mu R^{5/2}} \sqrt{\frac{6\Upsilon}{\mathrm{E}}}, \tag{S14}$$

where $\Upsilon$ is typically $9\,\mathrm{GPa}$ for silica [S7]. From the analyses made above, the mechanical limit of the spinning frequency can be given by:

$$\Omega_{\max} = \min\{\Omega_1, \Omega_2, \Omega_3\}. \tag{S15}$$

When operating at taper-resonator separations near $20\,\mathrm{nm}$ [S1], we find $\Omega_1 = 13.9\,\mathrm{kHz}$, $\Omega_2 = 893.6\,\mathrm{Hz}$, and $\Omega_3 = 133.3\,\mathrm{MHz}$, thereby the maximum value of angular velocity can be up to $893.6\,\mathrm{Hz}$. This indicates that the local deformation of the taper dominates the stability of our system and it is reasonable to set $\Omega = 600\,\mathrm{Hz}$ in the main text.

### B. Dissipative coupling between the counterpropagating modes

The other condition for realizing anti-$\mathcal{PT}$ symmetry is the dissipative coupling between the clockwise (CW) and counterclockwise (CCW) modes. According to a recent experiment on optical gyroscope [S8], the dissipative coupling can originate from the dissipative scattering induced by the fiber or any other dissipative scattering elements in the resonator. As given in Ref. [S8], in a standing-wave basis, the loss induced by the coupling of the fiber and the resonator can be expressed as

$$H_\mathrm{f} = \begin{pmatrix} a_1^\dagger & a_2^\dagger \end{pmatrix} \begin{pmatrix} -i\gamma_{a1} & 0 \\ 0 & -i\gamma_{a2} \end{pmatrix} \begin{pmatrix} a_1 \\ a_2 \end{pmatrix} = -i\gamma_{a1} a_1^\dagger a_1 - i\gamma_{a2} a_2^\dagger a_2, \tag{S16}$$

where $\gamma_{a1,a2}$ are the losses induced by the fiber for the two optical modes. Transforming this Hamiltonian in the traveling-mode basis with $a_{1,2} = (a_\mathrm{cw} \pm a_\mathrm{ccw})/\sqrt{2}$, we can get

$$\begin{aligned}H_\mathrm{f} = &-i(\gamma_{a1} + \gamma_{a2}) a_\mathrm{cw}^\dagger a_\mathrm{cw}/2 - i(\gamma_{a1} + \gamma_{a2}) a_\mathrm{ccw}^\dagger a_\mathrm{ccw}/2 \\ &+ i(\gamma_{a2} - \gamma_{a1}) a_\mathrm{cw}^\dagger a_\mathrm{ccw}/2 + i(\gamma_{a2} - \gamma_{a1}) a_\mathrm{ccw}^\dagger a_\mathrm{cw}/2.\end{aligned} \tag{S17}$$

Rewrite it in a matrix form

$$H_\mathrm{f} = \begin{pmatrix} a_\mathrm{cw}^\dagger & a_\mathrm{ccw}^\dagger \end{pmatrix} \begin{pmatrix} -i\gamma & i\kappa \\ i\kappa & -i\gamma \end{pmatrix} \begin{pmatrix} a_\mathrm{cw} \\ a_\mathrm{ccw} \end{pmatrix}, \tag{S18}$$

where $\gamma = (\gamma_{a1} + \gamma_{a2})/2$, $\kappa = (\gamma_{a2} - \gamma_{a1})/2$, and $\kappa$ is the effective dissipative coupling strength induced by the fiber. This dissipative coupling plays a crucial role in the realization of anti-$\mathcal{PT}$ symmetry in our scheme.

### C. Nanoparticle sensing

In previous experiments [S9, S10], nanoparticles falling onto the stationary resonator randomly were detected and counted. These nanoparticles can be deposited on the surface of the resonator via a nozzle [S9]. The complex optical mode coupling induced by a single particle, related to the overlap between the particle and the mode volume of the resonator, can be expressed as [S9]

$$\gamma_\mathrm{s} = \frac{2\pi^2 \alpha^2 f^2(\mathbf{r}) \omega_c}{3\lambda^3 V_c}, \quad g_\mathrm{s} = -\frac{\alpha f^2(\mathbf{r}) \omega_c}{2V_c}, \tag{S19}$$

where $\mathbf{r}$ is the particle position, $V_c$ is the mode volume of the resonator, $f(\mathbf{r})$ is the normalized mode distribution function, and $\lambda$ is the wavelength of the light. Here, $\alpha$ is the particle polarizability, which depends on the size

and refractive index of the particle. This complex modal coupling could induce mode splitting in the transmission spectrum which can be used to estimate the size of the nanoparticle in the detection process.

In our scheme based on a spinning resonator, we consider that the scatterers fall onto and stay on the surface of the resonator, rotating with it, similar to the particles on a stationary ring cavity [S9, S10]. A nanoparticle can be detected in such a system only when the splitting is resolved in the transmission spectrum, which requires $\gamma_s/g_s < 1$ [S9]. While in our scheme, the locations of the scatterers may be changed with the fast spinning of the resonator, thus affect the mode volume and the overlap of the particles, leading to the change in the values of $\gamma_s$ and $g_s$ according to Eq. (S19). But the ratio $\gamma_s/g_s = -4\pi^2\alpha/(3\lambda^3)$ will not be changed. Thus, we take the fixed value of $\gamma_s/g_s \sim 0.05$ in the numerical simulations of the perturbated system as in the experiments [S10, S11]. In experiments, the particle may fall through the mode volume of the resonator and not stay on the surface, resulting in a limited interaction time with the field, However, regardless of whether this time is very short or long, the resonator will feel and respond to it by exhibiting mode splitting [S12].

In this work, we compare in detail the different mode-splitting features for these two cases: the case with the scatter falling onto the anti-$\mathcal{PT}$ resonator and the other case with the scatterer falling onto the Hermitian spinning (HS) resonator [S1, S13]. We define the signal enhancement factor to evaluate the performance of these two sensors and show that the sensitivity is always enhanced by the anti-$\mathcal{PT}$ sensor (see Sec. S3). For both of the two sensors, if the particle is not stationary on the mode volume (particle diffuses on the surface) then we will observe that mode splitting will change. A particle located in a high intensity field in the mode volume will lead to a larger mode splitting than the same particle located in a low intensity field. If the particle is detached from the resonator due to spinning or any other reasons, then mode splitting will return back to the case when there is no particle. However, Sagnac effect and dissipative coupling themselves are not affected by the particle position. Thus the symmetry-broken enhancement in sensitivity originating from the interplay between Sagnac effect and dissipative coupling will not be affected by the position of the particle. We can conclude that the location of the particle will not change the fact that anti-$\mathcal{PT}$ sensor performs better than HS sensor. In addition, HS sensor performs better than a stationary-resonator sensor, i.e., diabolic point sensor, which has been revealed in Ref. [S13].

## S2. DERIVATION OF THE EFFECTIVE ANTI-$\mathcal{PT}$-SYMMETRIC HAMILTONIAN

We will compare the anti-$\mathcal{PT}$-symmetric Hamiltonian with the $\mathcal{PT}$-symmetric one in this section, and present the detailed derivation of the anti-$\mathcal{PT}$-symmetric Hamiltonian of the spinning resonator system. As proposed in Ref. [S14], non-Hermitian Hamiltonian can have entirely real eigenvalues if it is symmetric under combined $\mathcal{PT}$ operations. As its counterpart, the anti-$\mathcal{PT}$-symmetric Hamiltonian, which was first proposed in Ref. [S15], follows $\{\mathcal{PT}, H\} = 0$ mathematically. We express the two-mode Hamiltonian possessing $\mathcal{PT}$ symmetry and anti-$\mathcal{PT}$ symmetry in the matrix form as

$$H = \begin{pmatrix} a_1^\dagger & a_2^\dagger \end{pmatrix} M \begin{pmatrix} a_1 \\ a_2 \end{pmatrix}. \tag{S20}$$

A $\mathcal{PT}$-symmetric Hamiltonian is usually in a form of

$$M_{\text{PT}} = \begin{pmatrix} \omega & \kappa \\ \kappa^* & \omega^* \end{pmatrix}, \tag{S21}$$

TABLE S1. The summary of the coupled two-mode $\mathcal{PT}$-symmetric and anti-$\mathcal{PT}$-symmetric system. $\Delta_1$ and $\Delta_2$ are the detunings of the two optical modes. $\gamma_1$ and $\gamma_2$ are the optical loss or gain. $\kappa_1$ and $\kappa_2$ are the coupling strengths between the optical modes.

| Symmetry | | Detuning | Coupling Strength | Gain/Loss |
|---|---|---|---|---|
| $\mathcal{PT}$: | 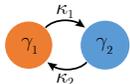 | $\Delta_1 = \Delta_2$ | $\kappa_1^* = \kappa_2$ | $\gamma_1 = -\gamma_2$ |
| Anti-$\mathcal{PT}$: | 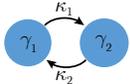 | $\Delta_1 = -\Delta_2$ | $\kappa_1^* = -\kappa_2$ | $\gamma_1 = \gamma_2$ |

where $\omega$ is the complex frequency and $\kappa$ is the complex coupling. Under the combined $\mathcal{PT}$ operations, it is invariant, i.e., $[\mathcal{PT}, H] = 0$. An anti-$\mathcal{PT}$-symmetric Hamiltonian is usually in the form of

$$M_{\mathrm{APT}} = \begin{pmatrix} \omega & \kappa \\ -\kappa^* & -\omega^* \end{pmatrix}. \tag{S22}$$

which will be mapped to its opposite under the combined $\mathcal{PT}$ operations, i.e., $\{\mathcal{PT}, H\} = 0$. We can see by multiplying $i$, the $\mathcal{PT}$-symmetric Hamiltonian can become anti-$\mathcal{PT}$ symmetric mathematically.

In Tab. S1, taking a two-mode coupled optical system as an example, we summarize the conditions that need to be satisfied in $\mathcal{PT}$-symmetric and anti-$\mathcal{PT}$-symmetric systems. For $\mathcal{PT}$-symmetric systems, one of the optical modes should be active with gain while the other mode has to be passive with equal loss. For the anti-$\mathcal{PT}$-symmetric Hamiltonian, the opposite detunings and the anti-Hermitian coupling should be met at the same time. Different from $\mathcal{PT}$ symmetry, the realization of anti-$\mathcal{PT}$ symmetry is independent on sophisticated gain-loss balance structure, see Fig. S2(a). The anti-$\mathcal{PT}$ symmetric system can be passive (with loss) [S16] or active (with gain) [S17].

In the following part, we will show how to construct the Hamiltonian in the form as shown in Eq. (S22). Inspired by the recent experiment on nonreciprocal light transmission with a spinning resonator [S1], we propose a scheme to realize anti-$\mathcal{PT}$ symmetry via a single linear optical resonator, which can support two counterpropagating modes, i.e., the CW and CCW modes. The Hamiltonian for such a system can be expressed as ($\hbar = 1$):

$$H_0 = (\omega_c - i\gamma_c)\left(a^\dagger_{\mathrm{cw}} a_{\mathrm{cw}} + a^\dagger_{\mathrm{ccw}} a_{\mathrm{ccw}}\right) + i\kappa \left(a^\dagger_{\mathrm{cw}} a_{\mathrm{ccw}} + a^\dagger_{\mathrm{ccw}} a_{\mathrm{cw}}\right), \tag{S23}$$

where $\omega_c = c/\lambda$ is the resonant frequency of the optical mode with $\lambda$ ($c$) being the wavelength (speed) of light, $\gamma_c = (\gamma_0 + \gamma_{\mathrm{ex}})/2$ is the total cavity loss, $\gamma_0 \equiv \omega_c/Q$ is the intrinsic loss of the cavity mode, $\gamma_{\mathrm{ex}}$ is the loss induced by the coupling between the fiber and the resonator, $\kappa$ is the dissipative coupling strength between the CW and the CCW modes, and $a_{\mathrm{cw}}(a_{\mathrm{ccw}})$ and $a^\dagger_{\mathrm{cw}}(a^\dagger_{\mathrm{ccw}})$ represent the annihilation and creation operators of the CW (CCW) cavity mode. This resonator is bidirectionally driven by two pumps with the same frequency $\omega_d$. The driving terms can be expressed as

$$H_{\mathrm{dr}} = i\varepsilon_d \left(a^\dagger_{\mathrm{cw}} e^{-i\omega_d t} - a_{\mathrm{cw}} e^{i\omega_d t}\right) + i\varepsilon_d \left(a^\dagger_{\mathrm{ccw}} e^{-i\omega_d t} - a_{\mathrm{ccw}} e^{i\omega_d t}\right), \tag{S24}$$

where $\varepsilon_d = \sqrt{\gamma_{\mathrm{ex}} P_d/\hbar\omega_d}$ is the driving amplitude and $P_d$ denotes the input power. Rewriting the Hamiltonian in the rotating frame with

$$U = \exp\left[-i\omega_d \left(a^\dagger_{\mathrm{cw}} a_{\mathrm{cw}} + a^\dagger_{\mathrm{ccw}} a_{\mathrm{ccw}}\right) t\right], \tag{S25}$$

we can derive the transformed Hamiltonian as

$$H' = i\frac{dU^\dagger}{dt}U + U^\dagger H U, \tag{S26}$$

where

$$H = H_0 + H_{\mathrm{dr}}. \tag{S27}$$

Then we get the transformed Hamiltonian as

$$H' = (\Delta_c - i\gamma_c)\left(a^\dagger_{\mathrm{cw}} a_{\mathrm{cw}} + a^\dagger_{\mathrm{ccw}} a_{\mathrm{ccw}}\right) + i\kappa \left(a^\dagger_{\mathrm{cw}} a_{\mathrm{ccw}} + a^\dagger_{\mathrm{ccw}} a_{\mathrm{cw}}\right) + i\varepsilon_d \left(a^\dagger_{\mathrm{cw}} - a_{\mathrm{cw}}\right) + i\varepsilon_d \left(a^\dagger_{\mathrm{ccw}} - a_{\mathrm{ccw}}\right), \tag{S28}$$

with driving detuning $\Delta_c = \omega_c - \omega_d$.

If this resonator rotates in the CCW direction at a speed of $\Omega$ (in unit of Hz) [S1], the Sagnac effect will result in two opposite frequency shifts in the CW and CCW modes [S21], i.e.,

$$\Delta_{\mathrm{sag}} = \pm\frac{nR\Omega\omega_c}{c}\left(1 - \frac{1}{n^2} - \frac{\lambda}{n}\frac{dn}{d\lambda}\right). \tag{S29}$$

TABLE S2. Experimentally feasible parameters used for numerical simulations [S1, S8, S18–S20]. Here $\lambda$ is the wavelength of the light, $Q$ is the quality factor of the resonator, $n$ is the refractive index of the resonator, $R$ is the radius of the resonator, $\kappa$ is the dissipative coupling, $P_d$ is the pump power, and $\gamma_{\mathrm{ex}}$ is the coupling loss.

| $\lambda$ | $Q$ | $n$ | $R$ | $\kappa$ | $P_d$ | $\gamma_{\mathrm{ex}}$ |
|---|---|---|---|---|---|---|
| 1550 nm | $\sim 1 \times 10^{11}$ | 1.44 | 50 $\mu$m | 8 kHz | 5 $\mu$W | $\gamma_0/2$ |

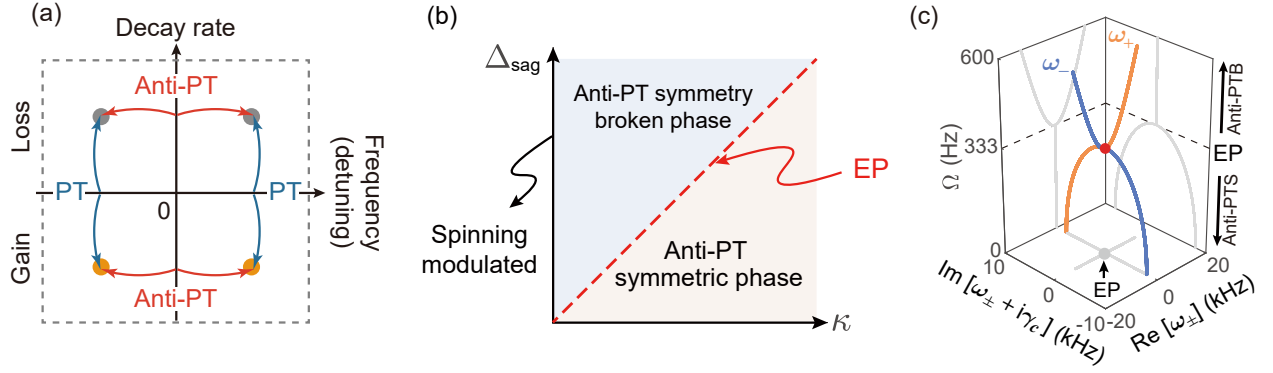

FIG. S2. (a) Differences between $\mathcal{PT}$ symmetry and anti-$\mathcal{PT}$ symmetry. (b) Schematic of the interplay between the dissipative coupling and the rotation-induced frequency shifts in anti-$\mathcal{PT}$-symmetric system. (c) The evolution of the real parts and imaginary parts of the eigenfrequencies as a function of the rotating speed $\Omega$. The coupling strength is $\kappa = 8\,\text{kHz}$ [S8] and the corresponding rotation rate at EP is $\Omega_{\text{EP}} \approx 333\,\text{Hz}$.

Hereafter, we replace $n_2$ with $n$ to denote the refractive index of the resonator just for convenience. The dispersion term $dn/d\lambda$, characterizing the relativistic origin of the Sagnac effect, is relatively small in typical materials ($\sim 1\%$) [S1]. Then the Hamiltonian becomes:

$$H' = (\Delta_+ - i\gamma_c)\, a_{\text{cw}}^\dagger a_{\text{cw}} + (\Delta_- - i\gamma_c)\, a_{\text{ccw}}^\dagger a_{\text{ccw}} + i\kappa \left(a_{\text{cw}}^\dagger a_{\text{ccw}} + a_{\text{ccw}}^\dagger a_{\text{cw}}\right) \\ + i\varepsilon_d \left(a_{\text{cw}}^\dagger - a_{\text{cw}}\right) + i\varepsilon_d \left(a_{\text{ccw}}^\dagger - a_{\text{ccw}}\right), \tag{S30}$$

with the detunings $\Delta_\pm = \Delta_c \pm \Delta_{\text{sag}}$.

By setting $\Delta_c = 0$, the matrix form of the Hamiltonian without the driving terms becomes:

$$H_0 = \begin{pmatrix} a_{\text{cw}}^\dagger & a_{\text{ccw}}^\dagger \end{pmatrix} M \begin{pmatrix} a_{\text{cw}} \\ a_{\text{ccw}} \end{pmatrix}, \tag{S31}$$

where

$$M = \begin{pmatrix} \Delta_{\text{sag}} - i\gamma_c & i\kappa \\ i\kappa & -\Delta_{\text{sag}} - i\gamma_c \end{pmatrix}. \tag{S32}$$

Obviously, this Hamiltonian is anti-$\mathcal{PT}$ symmetric under the combined $\mathcal{PT}$ operations:

$$\{\mathcal{PT},\ H_0\} = 0. \tag{S33}$$

So far we have constructed the anti-$\mathcal{PT}$-symmetric Hamiltonian using the opposite frequency shifts induced by mechanical rotation and the dissipative coupling induced by taper scattering. We will examine the non-Hermitian degeneracy known as exceptional point (EP) in this system by solving the following equation ($\mathbb{I}_2$ is the identity $2 \times 2$ matrix)

$$|M - \omega \mathbb{I}_2| = 0, \tag{S34}$$

i.e.,

$$\begin{vmatrix} \Delta_{\text{sag}} - i\gamma_c - \omega & i\kappa \\ i\kappa & -\Delta_{\text{sag}} - i\gamma_c - \omega \end{vmatrix} = 0. \tag{S35}$$

Then the eigenfrequencies of the anti-$\mathcal{PT}$-symmetric system can be obtained as:

$$\omega_\pm = -i\gamma_c \pm \sqrt{\Delta_{\text{sag}}^2 - \kappa^2}. \tag{S36}$$

The experimentally accessible parameters are listed in Tab. S2. We note that $Q$ has been improved to $10^{12}$ in experiment [S19], and the rotation speed $\Omega$, which is lower than $1\,\text{kHz}$ in numerical simulations, is also experimentally feasible [S1].

As shown in Fig. S2(b), the term under the square root in Eq. (S36) is negative when $\Delta_{\text{sag}} < \kappa$, leading to the difference of linewidth and enabling the system to enter the anti-$\mathcal{PT}$-symmetric (anti-PTS) phase. When increasing the rotation speed until $\Delta_{\text{sag}} > \kappa$, the term under the square root becomes positive, thus frequency difference occurs and the system is in anti-$\mathcal{PT}$-symmetry-broken (anti-PTB) phase. When $\Delta_{\text{sag}} = \kappa$, the two eigenfrequencies coalesce and the system is exactly at the EP. In Fig. S2(c), we plot the eigenfrequencies as a function of the rotation speed $\Omega$. Based on Eq. (S36), we fix the coupling strength at $\kappa = 8\,\text{kHz}$, and obtain the rotation rate at EP with the aid of Eq. (S29): $\Omega_{\text{EP}} \approx 333\,\text{Hz}$. In Fig. S2(c), when the rotation speed is below a critical value, i.e., $\Omega < \Omega_{\text{EP}}$, anti-$\mathcal{PT}$-symmetric phase appears with purely imaginary splitting. When $\Omega = \Omega_{\text{EP}}$, the coalescence occurs, and we can observe the broken phase of anti-$\mathcal{PT}$ symmetry with purely real splitting at $\Omega > \Omega_{\text{EP}}$.

## S3. NONRECIPROCAL LIGHT TRANSMISSION AND ULTRASENSITIVE NANOPARTICLE SENSING

### A. Symmetry-broken nonreciprocity

We consider a probe light of frequency $\omega_p$ to be incident from the left port of the system. The Hamiltonian can be expressed as

$$H' = \left(\Delta'_+ - i\gamma_c\right) a^\dagger_{\text{cw}} a_{\text{cw}} + \left(\Delta'_- - i\gamma_c\right) a^\dagger_{\text{ccw}} a_{\text{ccw}} + i\kappa \left(a^\dagger_{\text{cw}} a_{\text{ccw}} + a^\dagger_{\text{ccw}} a_{\text{cw}}\right) + i\epsilon_p \left(a^\dagger_{\text{cw}} - a_{\text{cw}}\right), \tag{S37}$$

where $\Delta'_\pm = \Delta_p \pm \Delta_{\text{sag}}$ with $\Delta_p = \omega_c - \omega_p$ being the probe detuning, $P_p$ denotes the probe power, and $\epsilon_p = \sqrt{\gamma_{\text{ex}} P_p / \hbar \omega_p}$ represents the amplitude of the probe light whose probe frequency is $\omega_p$. The equations of motion can be derived as

$$\begin{aligned}
\dot{a}_{\text{cw}} &= -i\left(\delta_p + \Delta_{\text{sag}}\right) a_{\text{cw}} + \kappa a_{\text{ccw}} + \epsilon_p, \\
\dot{a}_{\text{ccw}} &= -i\left(\delta_p - \Delta_{\text{sag}}\right) a_{\text{ccw}} + \kappa a_{\text{cw}},
\end{aligned} \tag{S38}$$

with $\delta_p = \Delta_p - i\gamma_c$. In the strong-driving regime, we can get the steady-state solutions by making $\dot{a}_{\text{cw}} = 0$ and $\dot{a}_{\text{ccw}} = 0$:

$$a_{\text{cw}} = \frac{-i\left(\delta_p - \Delta_{\text{sag}}\right) \epsilon_p}{\left(\delta_p + \Delta_{\text{sag}}\right)\left(\delta_p - \Delta_{\text{sag}}\right) + \kappa^2}. \tag{S39}$$

Applying the input-output relation [S22]

$$a^{\text{out}}_{\text{cw}} = a^{\text{in}}_{\text{cw}} - \sqrt{\gamma_{\text{ex}}}\, a_{\text{cw}}, \tag{S40}$$

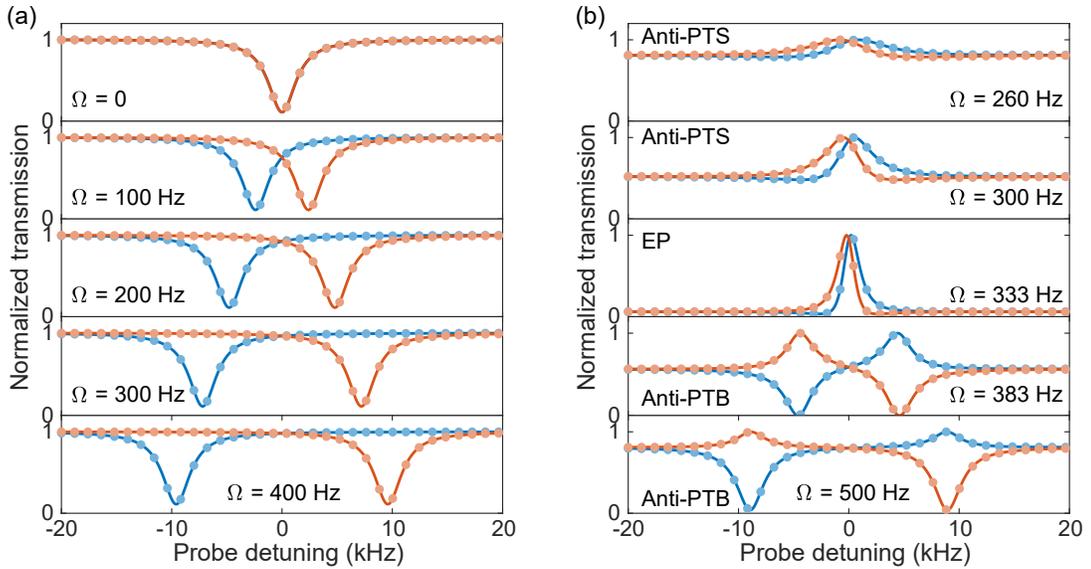

FIG. S3. (a) Normalized transmission rate of the HS system versus probe detuning. (b) Normalized transmission rate of the anti-$\mathcal{PT}$-symmetric system versus probe detuning. The blue curves represent the transmission in the CW direction while the orange curves stand for the transmission in the CCW direction. The parameters used here are the same as those used in Fig. S2.

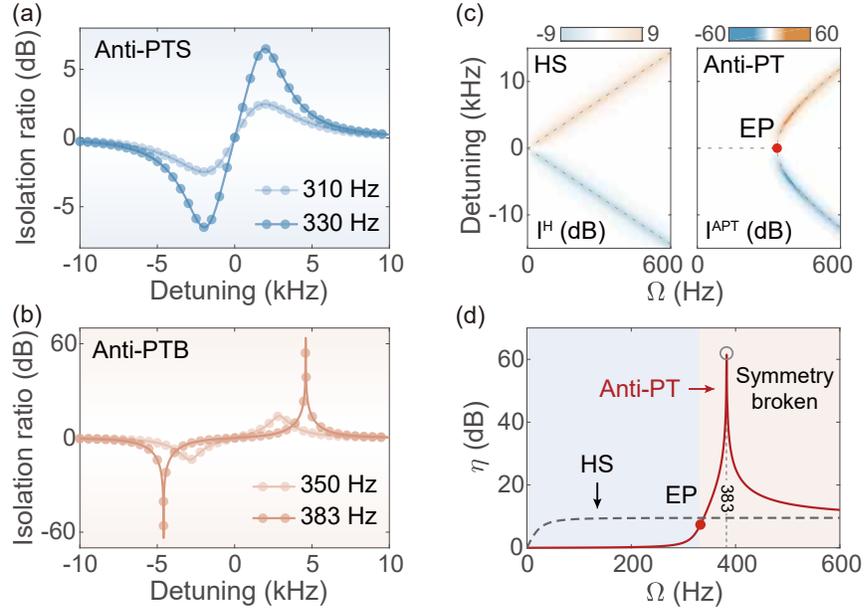

FIG. S4. Nonreciprocal transmission in the anti-$\mathcal{PT}$-symmetric system. (a) Isolation $I^{\mathrm{APT}}$ versus optical detuning $\Delta_p$ with different $\Omega$ in the anti-$\mathcal{PT}$-symmetric phase and (b) symmetry-broken phase. (c) Isolation ratios of Hermitian spinning (HS) resonator (left panel) [S1] and anti-$\mathcal{PT}$ system (right panel) versus $\Omega$ and $\Delta_p$. The gray dashed curves show the frequency evolutions of the two systems with respect to the rotation speed. (d) The maximum isolation $\eta$ as a function of $\Omega$ for anti-$\mathcal{PT}$ (red solid line) and HS (dark dashed line) systems. The parameters are the same as those in Fig. S2.

the transmission rate can be obtained as:

$$T_{\mathrm{cw}}^{\mathrm{APT}} = \left|1 - \frac{\gamma_{\mathrm{ex}} a_{\mathrm{cw}}}{\epsilon_p}\right|^2 = \left|1 + \frac{i\gamma_{\mathrm{ex}}(\delta_p - \Delta_{\mathrm{sag}})}{(\delta_p + \Delta_{\mathrm{sag}})(\delta_p - \Delta_{\mathrm{sag}}) + \kappa^2}\right|^2, \tag{S41}$$

where the superscript APT denotes anti-$\mathcal{PT}$-symmetric system. Similarly, when the probe is incident from the right, the solutions can be derived as

$$a_{\mathrm{ccw}} = \frac{-i(\delta_p + \Delta_{\mathrm{sag}})\epsilon_p}{(\delta_p + \Delta_{\mathrm{sag}})(\delta_p - \Delta_{\mathrm{sag}}) + \kappa^2}. \tag{S42}$$

According to the input-output relation

$$a_{\mathrm{ccw}}^{\mathrm{out}} = a_{\mathrm{ccw}}^{\mathrm{in}} - \sqrt{\gamma_{\mathrm{ex}}} a_{\mathrm{ccw}}, \tag{S43}$$

the transmission rate can be given by

$$T_{\mathrm{ccw}}^{\mathrm{APT}} = \left|1 - \frac{\gamma_{\mathrm{ex}} a_{\mathrm{ccw}}}{\epsilon_p}\right|^2 = \left|1 + \frac{i\gamma_{\mathrm{ex}}(\delta_p + \Delta_{\mathrm{sag}})}{(\delta_p + \Delta_{\mathrm{sag}})(\delta_p - \Delta_{\mathrm{sag}}) + \kappa^2}\right|^2. \tag{S44}$$

For comparison, we consider the light transmission in an HS system with single mode [S1]. The steady-state solutions with the probe light incident from the left and right are

$$a_{\mathrm{cw}} = \frac{-i\epsilon_p}{\delta_p + \Delta_{\mathrm{sag}}}, \quad a_{\mathrm{ccw}} = \frac{-i\epsilon_p}{\delta_p - \Delta_{\mathrm{sag}}}, \tag{S45}$$

respectively. Likewise, the corresponding transmissions are

$$T_{\mathrm{cw}}^{\mathrm{HS}} = \left|1 + \frac{i\gamma_{\mathrm{ex}}}{\delta_p + \Delta_{\mathrm{sag}}}\right|^2, \quad T_{\mathrm{ccw}}^{\mathrm{HS}} = \left|1 + \frac{i\gamma_{\mathrm{ex}}}{\delta_p - \Delta_{\mathrm{sag}}}\right|^2. \tag{S46}$$

In Fig. S3, the normalized transmission rate $T_{\mathrm{cw,ccw}}^s / \max[T_{\mathrm{cw,ccw}}^s]$, is plotted as a function of the probe detuning $\Delta_p$ at different rotation speed, where $s$ denotes the anti-$\mathcal{PT}$ (APT) system or the HS system. For the HS system,

starting from the stationary case ($\Omega = 0$), the countercirculating modes overlap, owing to their expected degeneracy. As predicted by Eq. (S29), increasing the mechanical rotation frequency $\Omega$ results in a linear opposing frequency shift for the countercirculating modes, i.e., nonreciprocal light transmission [S1]. For the anti-$\mathcal{PT}$-symmetric system, as shown in Fig. S3(b), there is no splitting in the transmission spectra in the symmetry-unbroken phase ($\Omega < \Omega_{\text{EP}}$), and nonreciprocal light transmission is unclear compared with the HS system, although the rotation speed has been increased over 200 Hz. This is owing to the fact that the rotation-induced Sagnac shift is smaller than the dissipative coupling and the transmission in two opposite directions is not greatly separated from each other [see Eqs. (S41) and (S44)]. Meanwhile, there is no dissipative coupling in HS system, thus the nonreciprocal transmission is only dependent on the Sagnac shift, as shown in Eq. (S46). The mode splittings for CW and CCW modes appear in the symmetry-broken phase ($\Omega > \Omega_{\text{EP}}$) by increasing the rotation speed, and the nonreciprocity becomes stronger than that in HS system.

To further confirm this picture, we study the isolation of this anti-$\mathcal{PT}$ system. And different from the main text, here we use the isolation ratio defined as $I^{\text{APT}} = 10\log_{10}\left(T_{\text{cw}}^{\text{APT}}/T_{\text{ccw}}^{\text{APT}}\right)$, and we find that the isolation ratio in symmetry-broken phase is much larger than that in anti-$\mathcal{PT}$-symmetric phase, as shown in Figs. S4(a) and S4(b). For comparison, the isolation ratio of HS resonator $I^{\text{HS}} = 10\log_{10}\left(T_{\text{cw}}^{\text{HS}}/T_{\text{ccw}}^{\text{HS}}\right)$ and the maxima of the isolation $\eta \equiv \max[I]$ for $\Delta_p \in [-15, 15]$ kHz with respect to $\Omega$ are revealed in Figs. S4(c) and S4(d), respectively. In HS system, the isolation becomes larger by increasing $\Omega$ due to the splitting of the counterpropagating modes induced by the Sagnac frequency shift; but it will be limited to 9.5 dB because of the fixed linewidths of the two modes [S1]. In contrast, the dissipative coupling combined with the Sagnac frequency shift can alter the linewidths and extrema of the transmission spectrum, thus the isolation rate can be improved to 61.8 dB in symmetry-broken phase. This is consistent with the results obtained in the main text. This anti-$\mathcal{PT}$-symmetry-broken enhanced nonreciprocity is related to the interplay between the linear synthetic angular momentum and dissipative backscattering. Different from the nonreciprocity reported in the $\mathcal{PT}$-symmetric systems [S23, S24], which relies on the nonlinear process, the nonreciprocal light transmission in this system is free of nonlinearity or gain-loss balanced structure.

### B. Anti-$\mathcal{PT}$ sensor

Apart from one-way control of transmission rates, in the following, we would like to present the potential of the anti-$\mathcal{PT}$-symmetric system for detection of nanoparticles [S9, S11, S13], which is highly desirable for widespread applications in, e.g., medical diagnosis and environmental monitoring [S9, S25, S26].

The Hamiltonian of this anti-$\mathcal{PT}$-symmetric system modified by nanoparticles can be expressed in a matrix form as [S9, S13, S27]:

$$M_N = \begin{pmatrix} \Delta_{\text{sag}} + \delta_N & i\kappa + C_N^- \\ i\kappa + C_N^+ & -\Delta_{\text{sag}} + \delta_N \end{pmatrix}, \tag{S47}$$

with

$$g_N = \sum_{i=1}^{N} g_{s,i}, \quad \gamma_N = \sum_{i=1}^{N} \gamma_{s,i}, \quad C_N^\pm = \sum_{i=1}^{N} (g_{s,i} - i\gamma_{s,i}) \exp(\pm i 2m\beta_i), \quad (i = 1, 2, 3, ..., N), \tag{S48}$$

where $\delta_N = g_N - i(\gamma_c + \gamma_N)$, $N$ is the total particle number, $m$ is the azimuthal mode number, $\beta_i$ is the angular position of the $i$-th nanoparticle, and $g_{s,i}$ ($\gamma_{s,i}$) is the coupling strength (the loss rate) induced by the $i$-th nanoparticle. According to $|M_N - \omega \mathbb{I}_2| = 0$, the eigenfrequencies of this perturbed system can be given by:

$$\omega_N^\pm = \delta_N \pm \sqrt{\Delta_{\text{sag}}^2 + (i\kappa + C_N^-)(i\kappa + C_N^+)}. \tag{S49}$$

The corresponding frequency difference is

$$\Delta\omega_N^{\text{APT}} = \omega_N^+ - \omega_N^- = 2\sqrt{\Delta_{\text{sag}}^2 + (i\kappa + C_N^-)(i\kappa + C_N^+)}. \tag{S50}$$

To explore EP assisted sensitive sensing, we compare the anti-$\mathcal{PT}$ sensor with HS sensor [S13]. In the presence of perturbations, the eigenfrequencies for HS sensor can be written as

$$\omega_N^\pm = \delta_N \pm \sqrt{\Delta_{\text{sag}}^2 + C_N^- C_N^+}, \tag{S51}$$

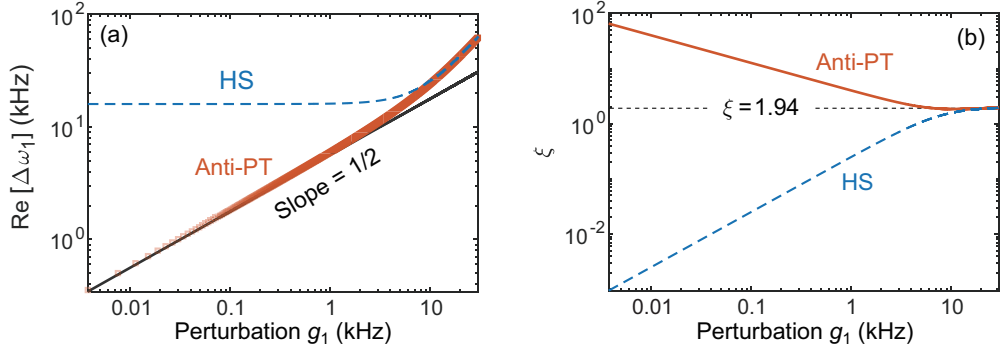

FIG. S5. Anti-$\mathcal{PT}$-symmetric ultrasensitive nanoparticle sensing. (a) The real part of the complex frequency splitting $\text{Re}\left[\Delta\omega_1\right]$ versus the perturbation $g_1$ corresponding to the anti-$\mathcal{PT}$ and HS sensors. (b) The signal enhancement factor as a function of the perturbation $g_1$. The rotation speed of these two sensors is fixed at $\Omega_{\text{EP}} \approx 333$ Hz. $\gamma_1/g_1$ is set to be 0.05 as in the experiments [S10, S11]. The other parameters used here are the same as those in Fig. S2.

thus yield the frequency difference as follows:

$$\Delta\omega_N^{\text{HS}} = 2\sqrt{\Delta_{\text{sag}}^2 + C_N^- C_N^+}. \tag{S52}$$

It should be noted that the coupling between CW and CCW modes in HS sensor is only induced by nanoparticles.

By setting $N = 1$, we consider the simplest case that only a single nanoparticle is deposited on the resonator with $g_1 = g_{s,1}$ and $\gamma_1 = \gamma_{s,1}$. In this case, angular position $\beta_1$ is set to be 0, leading to $C_1^+ = C_1^- = g_{s,1} - i\gamma_{s,1} = g_1 - i\gamma_1$. Here we have restricted our discussion to $J = g_1 - i\gamma_1$ for convenience, then the perturbed eigenfrequencies can be described as

$$\omega_1^\pm = J - i\gamma_c \pm \sqrt{\Delta_{\text{sag}}^2 - (\kappa - iJ)^2}, \tag{S53}$$

with the corresponding frequency difference:

$$\Delta\omega_1^{\text{APT}} = \omega_1^+ - \omega_1^- = 2\sqrt{\Delta_{\text{sag}}^2 - (\kappa - iJ)^2}. \tag{S54}$$

Likewise, we can write perturbed frequencies and their difference for HS sensor as follows:

$$\omega_1^\pm = J - i\gamma_c \pm \sqrt{\Delta_{\text{sag}}^2 + J^2}, \quad \Delta\omega_1^{\text{HS}} = 2\sqrt{\Delta_{\text{sag}}^2 + J^2}. \tag{S55}$$

In Fig. S5(a), the logarithmic behaviour of the real part of the complex frequency splitting is shown to highlight the sensitivity enhancement of anti-$\mathcal{PT}$ sensor. For the same minuscule perturbation, anti-$\mathcal{PT}$ sensor at EP performs better than HS sensor, which shows no strong dependence on perturbation. However, their behaviours are similar for large disturbance. This can be explained with the perturbation theory. The complex frequency splitting at EP ($\Delta_{\text{sag}} = \kappa$) is expected to approximately follow

$$\Delta\omega_1^{\text{APT}} = 2\sqrt{2i\kappa}J^{1/2}, \tag{S56}$$

when $J$ is much smaller than $\kappa$, indicating that perturbations experience an enhancement of the form $J^{1/2}$. For larger $J$, the slope of the splitting will be slightly larger than $1/2$, because in this case Eq. (S54) cannot be simplified to Eq. (S56), and the higher order terms should be taken into consideration:

$$\Delta\omega_1^{\text{APT}} = 2\sqrt{2i\kappa}J^{1/2} - \frac{i\sqrt{2i\kappa}}{2\kappa}J^{3/2} + \cdots. \tag{S57}$$

For HS sensor, the complex frequency splitting

$$\Delta\omega_1^{\text{HS}} = 2\Delta_{\text{sag}} + \frac{J^2}{\Delta_{\text{sag}}} - \frac{J^4}{4\Delta_{\text{sag}}^3} + \cdots, \tag{S58}$$

is proportional to the square of the perturbation $J^2$ at least. Thus, HS sensor will not show strong dependence on small perturbation.

A signal enhancement factor defined as

$$\xi = \left|\frac{\partial \Delta \omega_1^{\text{APT}}}{\partial g_1}\right| = \left|\frac{2(\kappa - \gamma_1 - ig_1)(\gamma_1/g_1 + i)}{\sqrt{\Delta_{\text{sag}}^2 - (\kappa - \gamma_1 - ig_1)^2}}\right|, \quad (S59)$$

is introduced to intuitively evaluate the performance of anti-$\mathcal{PT}$ sensor. Similarly, the signal enhancement factor for HS sensor can be given by

$$\xi = \left|\frac{\partial \Delta \omega_1^{\text{HS}}}{\partial g_1}\right| = \left|\frac{2(-\gamma_1 - ig_1)(\gamma_1/g_1 + i)}{\sqrt{\Delta_{\text{sag}}^2 + (-i\gamma_1 + g_1)^2}}\right|. \quad (S60)$$

The dependence of signal enhancement factor $\xi$ on perturbations is shown in Fig. S5(b), from which we can see in the vicinity of EP, larger sensitivity can be obtained for smaller perturbations in anti-$\mathcal{PT}$ sensor. While in HS sensor, the signal enhancement factor is small for minuscule perturbations. When the perturbation $J$ is much larger than $\kappa$, as aforementioned, the signal enhancement factor of these two sensors will be roughly equal ($\sim 2$, as illustrated in the figure) for the fact that the higher order terms in Eqs. (S57) and (S58) cannot be neglected. One of the potential applications of this anti-$\mathcal{PT}$-symmetric system, i.e., anti-$\mathcal{PT}$ based sensing and its enhancement to the sensitivity are carefully confirmed, which could open a new path towards engineering compacted ultrasensitive sensors for detections of nanoscale objects.

Alternatively, we can evaluate the performance of these two sensors from the optical fields in one of the counterpropagating modes, e.g., the CW mode. The effective Hamiltonian of the anti-$\mathcal{PT}$-symmetric system with nanoparticles can be written as

$$H_N = (\Delta'_+ + \delta_N) a_{\text{cw}}^\dagger a_{\text{cw}} + (\Delta'_- + \delta_N) a_{\text{ccw}}^\dagger a_{\text{ccw}} + (i\kappa + C_N^-) a_{\text{cw}}^\dagger a_{\text{ccw}} + (i\kappa + C_N^+) a_{\text{ccw}}^\dagger a_{\text{cw}} + i\epsilon_p (a_{\text{cw}}^\dagger - a_{\text{cw}}). \quad (S61)$$

The equations of motion can be derived as

$$\begin{aligned}\dot{a}_{\text{cw}} &= -i(\Delta'_+ + \delta_N) a_{\text{cw}} - i(i\kappa + C_N^-) a_{\text{ccw}} + \epsilon_p, \\ \dot{a}_{\text{ccw}} &= -i(\Delta'_- + \delta_N) a_{\text{ccw}} - i(i\kappa + C_N^+) a_{\text{cw}}.\end{aligned} \quad (S62)$$

The steady-state solutions can be obtained by making $\dot{a}_{\text{cw}} = 0$ and $\dot{a}_{\text{ccw}} = 0$, i.e.,

$$\begin{aligned}0 &= -i(\Delta'_+ + \delta_N) a_{\text{cw}} - i(i\kappa + C_N^-) a_{\text{ccw}} + \epsilon_p, \\ 0 &= -i(\Delta'_- + \delta_N) a_{\text{ccw}} - i(i\kappa + C_N^+) a_{\text{cw}}.\end{aligned} \quad (S63)$$

Finally, we get

$$a_{\text{cw}} = \frac{-i(\Delta'_- + \delta_N)\epsilon_p}{(\Delta'_+ + \delta_N)(\Delta'_- + \delta_N) - (i\kappa + C_N^-)(i\kappa + C_N^+)}. \quad (S64)$$

Here we consider only one single nanoparticle is deposited on the resonator ($N = 1$), then the steady-state solution becomes

$$a_{\text{cw}} = \frac{-i(\Delta'_- + \delta_1)\epsilon_p}{(\Delta'_+ + \delta_1)(\Delta'_- + \delta_1) - (i\kappa + g_1 - i\gamma_1)^2}. \quad (S65)$$

Applying the input-output relation in Eq. (S40), we can get the transmission rate as

$$T_1^{\text{APT}} = \left|1 - \frac{\gamma_{\text{ex}} a_{\text{cw}}}{\epsilon_p}\right|^2. \quad (S66)$$

For HS sensor, the steady-state solution is

$$a_{\text{cw}} = \frac{-i(\Delta'_- + \delta_1)\epsilon_p}{(\Delta'_+ + \delta_1)(\Delta'_- + \delta_1) - (g_1 - i\gamma_1)^2}. \quad (S67)$$

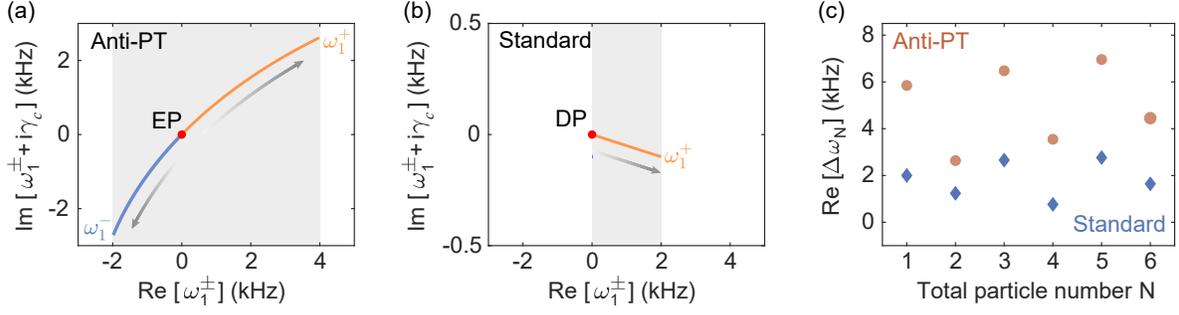

FIG. S6. Anti-$\mathcal{PT}$ sensor compared with standard sensor. (a,b) The evolution of the eigenfrequencies in the complex plane with the perturbation $g_1$ varying from 0 to 1 kHz. (a) shows the complex evolution corresponding to anti-$\mathcal{PT}$ sensor originally at EP. (b) shows the case of standard sensor. The perturbation increases along the arrow direction. The rotation speed is fixed corresponding to the EP ($\Omega_{\mathrm{EP}} \approx 333$ Hz) of the system without nanoparticle. $\gamma_1/g_1$ is set to be 0.05 as in the experiments [S10, S11]. (c) The frequency splitting of anti-$\mathcal{PT}$ sensor in the presence of multiparticles. The circles represent anti-$\mathcal{PT}$ sensor while the diamonds represent the standard sensor. The other parameters used here are the same as those in Fig. S2.

The corresponding transmission is

$$T_1^{\mathrm{HS}} = \left| 1 - \frac{\gamma_{\mathrm{ex}} a_{\mathrm{cw}}}{\epsilon_p} \right|^2. \tag{S68}$$

To compare the performance of the these two sensors, we introduce an enhancement factor

$$\chi = \mathcal{V}_{\mathrm{APT}}/\mathcal{V}_{\mathrm{HS}}, \tag{S69}$$

with the relative variation ratio defined as $\mathcal{V}_{\mathrm{APT}} = T_1^{\mathrm{APT}}/T_0^{\mathrm{APT}}$ and $\mathcal{V}_{\mathrm{HS}} = T_1^{\mathrm{HS}}/T_0^{\mathrm{HS}}$. Here $T_0^{\mathrm{APT}}$ ($T_0^{\mathrm{HS}}$) is the transmission for anti-$\mathcal{PT}$ (HS) sensor without nanoparticle (i.e., $g_1 = 0$ and $\gamma_1 = 0$). The probe detuning is chosen to be $\Delta_p = \mathrm{Re}\,[\omega_-]$ (see the main text).

Finally, we show that this anti-$\mathcal{PT}$ sensor still performs better than standard sensor system with diabolic point (DP). The perturbed eigenfrequencies and frequency splitting for standard sensor are

$$\omega_N^\pm = \delta_N \pm \sqrt{C_N^- C_N^+}, \ \Delta\omega_N^{\mathrm{D}} = 2\sqrt{C_N^- C_N^+}. \tag{S70}$$

For $N = 1$, the perturbed eigenfrequencies become $\omega_1^+ = -i\gamma_c + 2J$ and $\omega_1^- = -i\gamma_c$, and the frequency splitting is $\Delta\omega_1^{\mathrm{D}} = 2J$.

Figures S6(a) and S6(b) show the evolution of the eigenfrequencies corresponding to anti-$\mathcal{PT}$ sensor and standard sensor in the complex plane with $g_1$ varying from 0 to 1 kHz. The width of the shadow regions shown in Figs. S6(a) and S6(b) indicates the frequency splittings $\mathrm{Re}\,[\Delta\omega_1^s]$ induced by the same perturbation for these two sensors. Evidently, anti-$\mathcal{PT}$ sensor undergoes more splitting than the standard sensor. This is still true for multiparticle detection, as shown in Fig. S6(c). The values of $g_{\mathrm{s},i}$, $\beta_i$, and $\gamma_{\mathrm{s},i}$ are random in this figure. The azimuthal mode number is set to be $m = 4$ [S10].


\clearpage

\end{document}